\documentclass{PoS}

\usepackage{overpic}

\newcommand{\mpt}{\ensuremath{/\!\!\!\!{p}_{\rm T}}}

\newcommand{\fb}{\ensuremath{{\rm fb}^{-1}}}

\newcommand{\pt}{\ensuremath{p_{\rm T}}}

\newcommand{\etal}{{\em et. al.}}

\newcommand{\GeV}{\ensuremath{\textnormal{GeV}}}

\newcommand{\dzero}{D\O}

\newcommand{\ttbar}{\ensuremath{t\bar{t}}}

\newcommand{\etmissx}{\ensuremath{E \kern-0.6em\slash_{\rm x}}}
\newcommand{\etmissy}{\ensuremath{E \kern-0.6em\slash_{\rm y}}}

\newcommand{\dm}{\ensuremath{\Delta m}}
\newcommand{\mtop}{\ensuremath{m_{\rm top}}}
\newcommand{\mt}{\ensuremath{m_t}}
\newcommand{\mtb}{\ensuremath{m_{\bar t}}}
\newcommand{\ejets}{\ensuremath{e+{\rm jets}}}
\newcommand{\mujets}{\ensuremath{\mu+{\rm jets}}}
\newcommand{\ljets}{\ensuremath{\ell+{\rm jets}}}

\newcommand{\sigtt}{\ensuremath{\sigma_{t\bar t}}}
\newcommand{\mpole}{\ensuremath{m_{\rm top}^{\rm pole}}}
\newcommand{\mmsbar}{\ensuremath{m_{\rm top}^{\overline{\rm MS}}}}
\newcommand{\mmc}{\ensuremath{m_{\rm top}^{\rm MC}}}
\newcommand{\prob}{\ensuremath{\mathcal{P}}}
\newcommand{\psig}{\ensuremath{\mathcal{P}_{\rm sig}}}
\newcommand{\kjes}{\ensuremath{k_{\rm JES}}}

\title{Measurements of the top quark mass at D0}

\ShortTitle{Measurements of the top quark mass at D0}

\author{\speaker{Oleg Brandt} on behalf of the D0 collaboration\\%
        II. Physikalisches Institut, University of G\"ottingen\\
        E-mail: \email{obrandt@fnal.gov}}

\abstract{The mass of the top quark is a fundamental parameter of the standard model (SM) and has to be determined experimentally. The D0 experiment at the Fermilab Tevatron proton-antiproton collider with a centre-of-mass energy of 1.96~TeV has measured the top quark in various channels. In this talk, I present the most recent measurements of the top quark mass in the dilepton and lepton$+$jets channels with up to 5.3~\fb\ as well as their combination, and give an outlook on the final, most precise measurement of the top quark mass at D0.}

\FullConference{36th International Conference on High Energy Physics\\
     4-11 July 2012\\
     Melbourne, Australia}

\begin{document}

\section{Introduction}
The pair-production of the top quark was discovered in 1995 by the CDF and D0 experiments~\cite{bib:topdiscovery} at the Fermilab Tevatron proton-antiproton collider. Observation of the electroweak production of single top quarks was presented only two years ago~\cite{bib:singletop}. The large top quark mass and the resulting Yukawa coupling of almost unity indicates that the top quark could play a crucial role in electroweak symmetry breaking. Precise measurements of the properties of the top quark provide a crucial test of the consistency of the SM and could hint at physics beyond the SM. 

\begin{figure}
\centering
\begin{overpic}[height=0.335\textwidth]{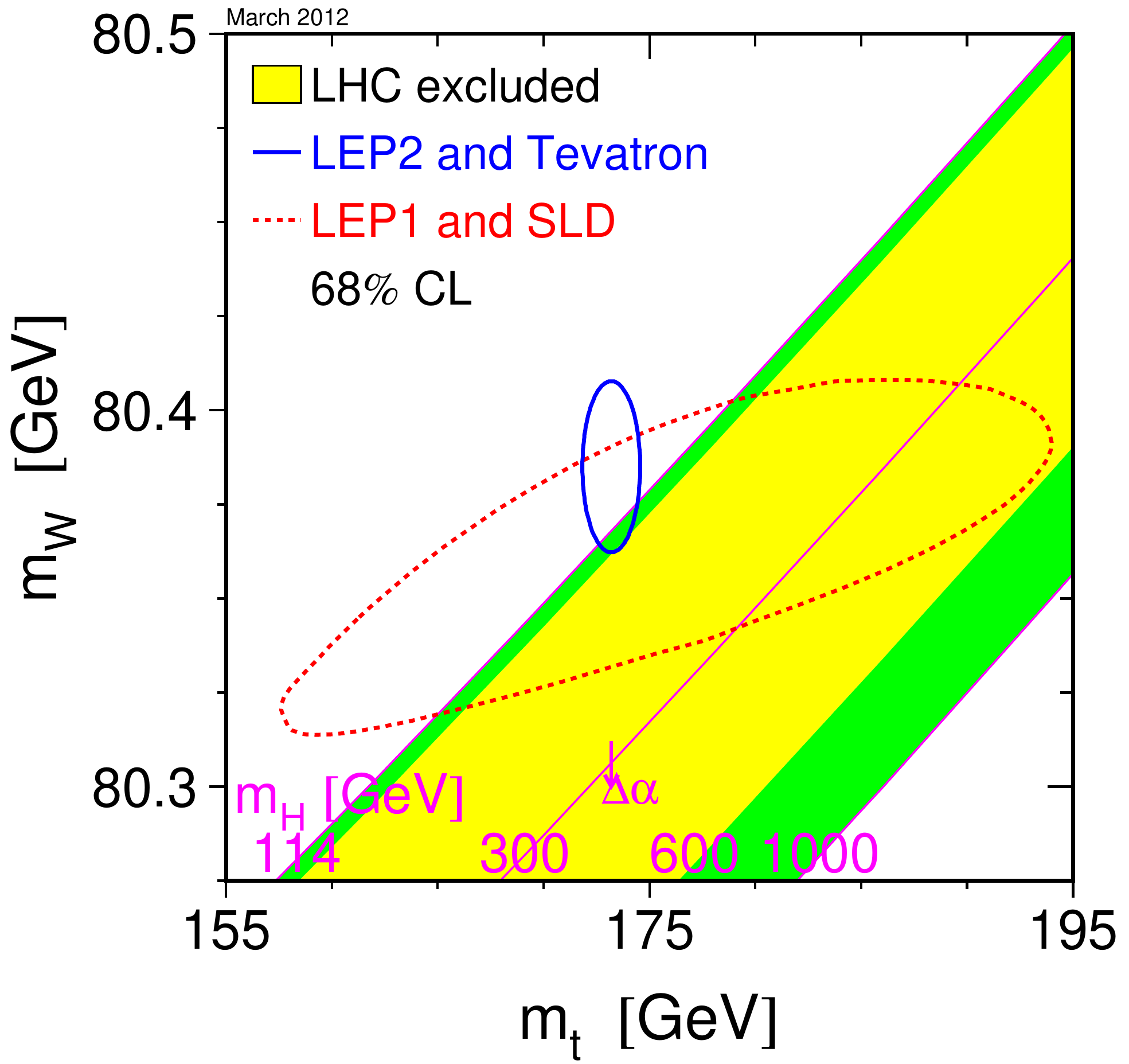}
\put(-1,2){(a)}
\end{overpic}
\qquad
\begin{overpic}[height=0.33\textwidth]{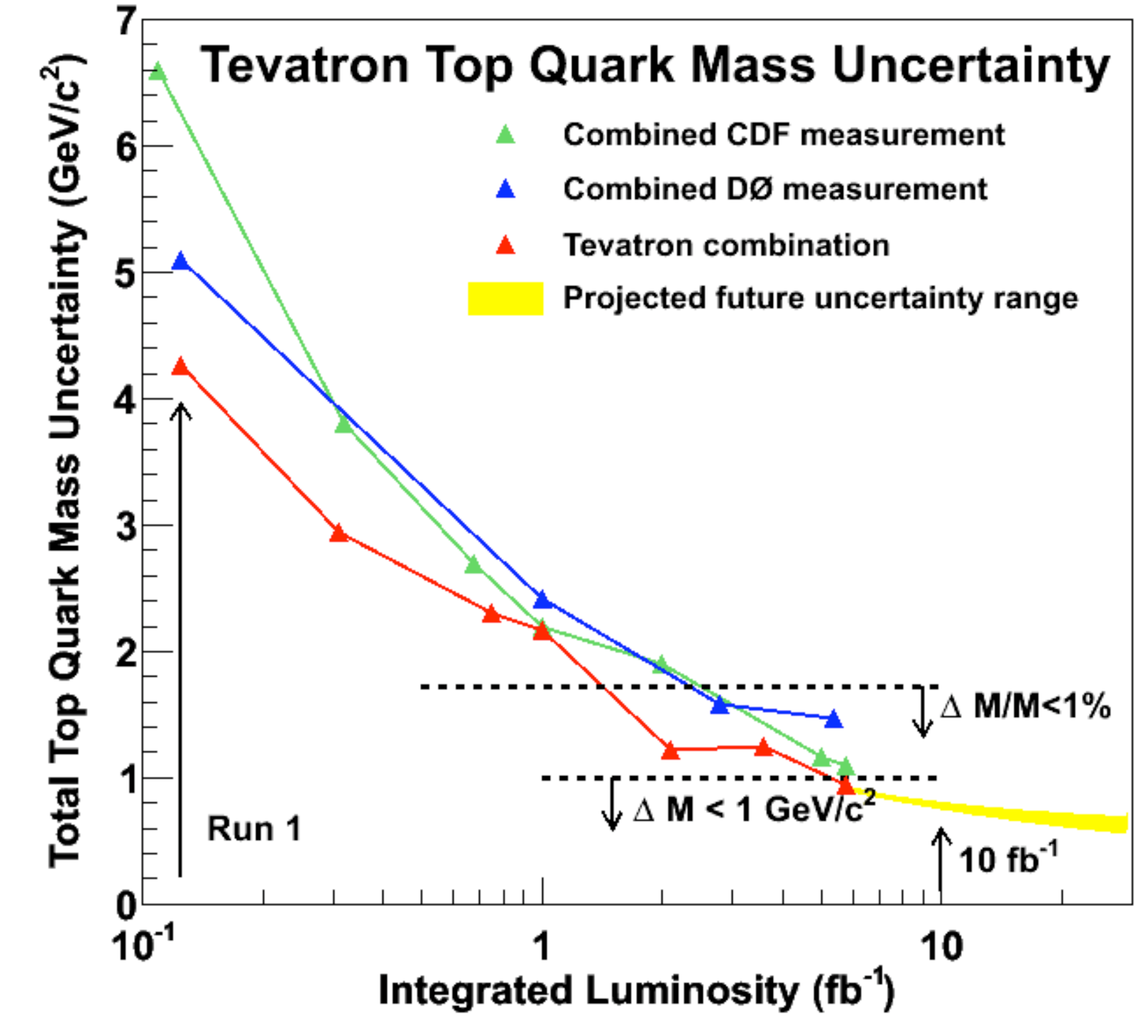}
\put(-1,2){(b)}
\end{overpic}
\caption{
\label{fig:mhiggs}
{\bf(a)} The constraint on mass of the SM Higgs boson from direct $\mtop$ and $m_W$ measurements in the $\mtop$,$m_W$ plane. The blue ellipsis indicates the 68\% CL contour. {\bf(b)} The anticipated precision on \mtop\ measurements at D0 and the Tevatron combination versus integrated luminosity.
}
\end{figure}

In the following, we review measurements of the top quark mass at the D0 experiment, which is a fundamental parameter of the SM. Its precise knowledge, together with the mass of the $W$~boson ($m_W$), provides an important constraint on the mass of the SM Higgs boson. This is illustrated in the \mtop,$m_W$ plane in Fig.~\ref{fig:mhiggs}, which includes the recent, most precise measurements of $m_W$~\cite{bib:wmasstalk}. Measurements of properties of the top quark other than \mtop\ are reviewed in Ref.~\cite{bib:proptalk}. The full listing of top quark measurements at D0 can be found in~Refs.~\cite{bib:topresd0}.

At the Tevatron, top quarks are mostly produced in pairs via the strong interaction. 
By the end of Tevatron operation, about 10.7 fb$^{-1}$ of integrated luminosity were recorded by D\O, which corresponds to about 80k produced $\ttbar$ pairs. In the framework of the SM, the top quark decays to a $W$~boson and a $b$~quark nearly 100\% of the time, resulting in a $W^+W^-b\bar b$ final state from top quark pair production. 
Thus, $\ttbar$ events are classified according to the $W$ boson decay channels as ``dileptonic'', ``all--jets'', or ``lepton+jets''. More details on the channels and their experimental challenges can be found in Ref.~\cite{bib:xsec}, while the electroweak production of single top quarks is reviewed in Ref.~\cite{bib:singletoptalk}.

\section{Direct measurements of the top quark mass in \ljets\ final states}
D0's most precise measurement of \mtop\ is performed in $\ell+4{\,\rm jets}$ final state using the so-called {\em matrix element} (ME) method in 3.6~\fb\ of data~\cite{bib:mtoplj_d0}. This technique was pioneered by \dzero\ in Run~I of the Tevatron~\cite{bib:topmassd0nature}, and it calculates the probability that a given event, characterised by a set of measured observables $x$, comes from the \ttbar\ production given an \mtop\ hypothesis, or from a background process: 
$\prob_{\rm evt}(x,\mtop)\propto f\prob_{\rm sig}(x,\mtop)+(1-f)\prob_{\rm bgr}(x)$.
The dependence on $\mtop$ is explicitly introduced by calculating $\psig$ using the differential cross section ${\rm d}\sigma(y,\mtop)\propto|\mathcal{M}_{\ttbar}|^2(\mtop)$, where $\mathcal{M}_{\ttbar}$ is the leading order (LO) matrix element for \ttbar\ production:
\[
\psig(x,\mtop,\kjes) = \frac1{\sigtt^{\rm observed}}\cdot\int W(x,y,\kjes)~{\rm d}\sigma(y,\mtop)\,.
\]
Since ${\rm d}\sigma(y,\mtop)$ is defined for a set of parton-level observables $y$, the transfer function $W(x,y,\kjes)$ is used to map them to the reconstruction-level set $x$. This accounts for detector resolutions and acceptance cuts, and introduces explicitly the dependence on the jet energy scale (JES) via an overall scaling factor \kjes. The uncertainty on the JES, which is almost fully correlated with \mtop, is around 2\% or larger. Therefore, an {\em in situ} calibration is performed by requiring that the mass of the dijet system assigned to the parton pair from the hadronically decaying $W$ boson be $m_{jj}=80.4~\GeV$. Thus, $\mtop$ and $\kjes$ are extracted simultaneously. This reduces the uncertainty from the JES to about 0.5\%, decreasing with the number of selected $\ttbar$ events. The measurement is performed in events with four jets, resulting in 24 possible jet-parton assignments. All 24 are summed over, weighted according to the consistency of a given assignment with the $b$-tagging information. $\prob_{\rm bgr}$ is calculated using the VECBOS matrix element for $W+4$~jets production. Generally, the ME technique offers a superior statistical sensitivity as it uses the full topological and kinematic information in the event in form of 4-vectors. The drawback of this method is the high computational demand.

D0 measures $\mtop=174.9 \pm 0.8~({\rm stat}) \pm 0.8~({\rm JES}) \pm 1.0~({\rm syst})~\GeV$, corresponding to a relative uncertainty of 0.9\%. The dominant systematic uncertainties are from modeling of underlying event activity and hadronisation, as well as the colour reconnection effects. On the detector modeling side, diffential uncertainties on the JES which are compatible with the overall $\kjes$ value from {\em in situ} calibration, and the difference between the JES for light and b-quark jets are dominant. This picture is representative for all \mtop\ measurements in \ljets\ final states shown here.

\section{Measurement of the top quark mass in dilepton final states using the matrix element technique}
We measure \mtop\ in dilepton final states with 5.4~\fb\ of data~\cite{bib:mtopllme_d0} using the ME technique similar to the one used in the \ljets\ channel. Leaving \mtop\ as a free parameter, dilepton final states are kinematically underconstrained by two degrees of freedom, and a prior is assumed for the transverse momentum distribution of the \ttbar\ system. The neutrino momenta are integrated over. The dominant background contribution comes from $Z+{\rm 2~jets}$ events, and the corresponding LO ME is used to represent the background. No {\em in situ} calibration is possible in dilepton final states, making the JES and the JES of $b$ quark jets the dominant source of systematic uncertainty. After calibration, a top quark mass of $174.0~ \pm 1.8~({\rm stat}) \pm 2.4~({\rm syst})~\GeV$ is found.

\section{Measurement of the top quark mass in dilepton final states using the neutrino weighting technique}
The world's most precise measurement of \mtop\ in dilepton final states is performed by D0 using 4.7~\fb\ of data~\cite{bib:mtopll_d0}. To account for the kinematically underconstrained degree of freedom, the so-called neutrino weighting algorithm is applied for kinematic reconstruction. It postulates distributions in rapidities of the neutrino and the antineutrino, and calculates a weight, which depends on the consistency of the reconstructed $\vec\pt^{\,\nu\bar\nu}\equiv\vec\pt^{\,\nu}+\vec\pt^{\,\bar\nu}$ with the measured missing transverse momentum $\mpt$ vector, versus \mtop. D0 uses the first and second moment of this weight distribution to define templates and extract \mtop. To reduce the systematic uncertainty, the {\em in situ} JES calibration in \ljets\ final states derived in Ref.~\cite{bib:mtoplj_d0} is applied, accounting for differences in jet multiplicity, luminosity, and detector ageing. After calibration and all corrections, $\mtop=174.0~ \pm 2.4~({\rm stat}) \pm 1.4~({\rm syst})~\GeV$ is found.

\section{Measurement of \mtop\ from the \ttbar\ production cross-section}
The $\ttbar$ production cross section (\sigtt) is correlated to \mtop. This can be used to extract \mtop\ by comparing the measured \sigtt\ to the most complete to--date, fully inclusive theoretical predictions, assuming the validity of the SM. Such calcualtions offer the advantage of using mass definitions in well-defined renormalisation schemes like $\mmsbar$ or $\mpole$. In contrast, the main methods using kinematic fits utilise the mass definition in MC generators $\mmc$, which cannot be translated into $\mmsbar$ or $\mpole$ in a straightforward way.
D0 uses 5.3~\fb\ of data to measure $\sigtt$ and extracts \mtop~\cite{bib:mtoppole_d0} using theoretical calculations for $\sigtt$ like the next-to-leading order (NLO) calculation with next-to-leading logarithmic (NLL) terms resummed to all orders~\cite{bib:signlo}, an approximate NNLO calculation~\cite{bib:signnlo}, and others. For this, a correction is derived to account for the weak dependence of $\sigtt$ on $\mmc$. The results for \mpole are presented in Fig.~\ref{fig:mpole}, and can be summarised as follows: $\mpole=163.0^{+5.1}_{-4.6}~\GeV$ and $\mpole=167.5^{+5.2}_{-4.7}~\GeV$ for Ref.~\cite{bib:signlo} and~\cite{bib:signnlo}, respectively. The effect from interpreting $\mmc$ as $\mmsbar$ or $\mpole$ is found to be about 3~\GeV.

\section{Measurements of the mass difference between the $t$ and $\bar t$ quarks}
\begin{figure}
\centering
\begin{overpic}[height=0.3\textwidth]{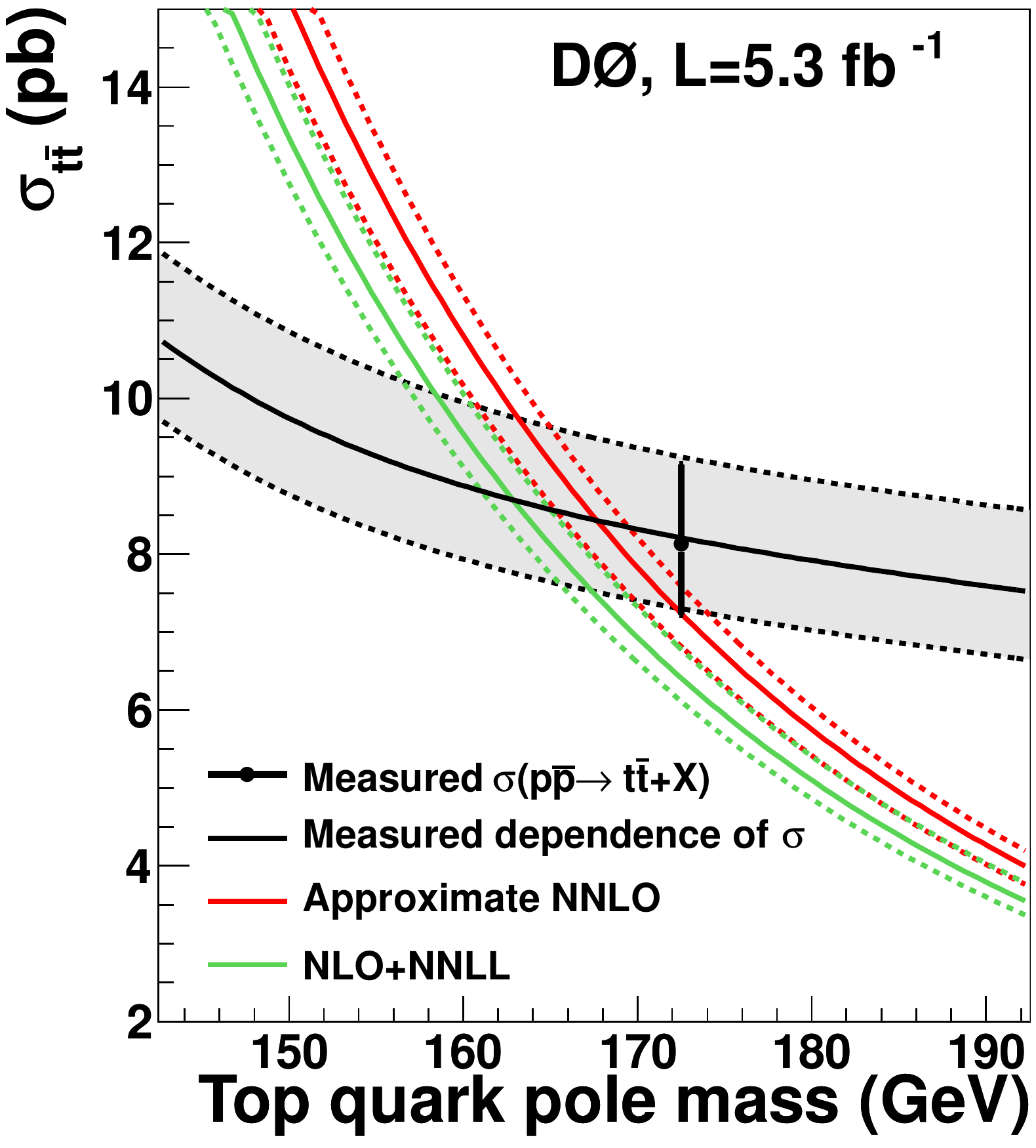}
\put(-1,2){(a)}
\end{overpic}
\qquad
\begin{overpic}[height=0.3\textwidth]{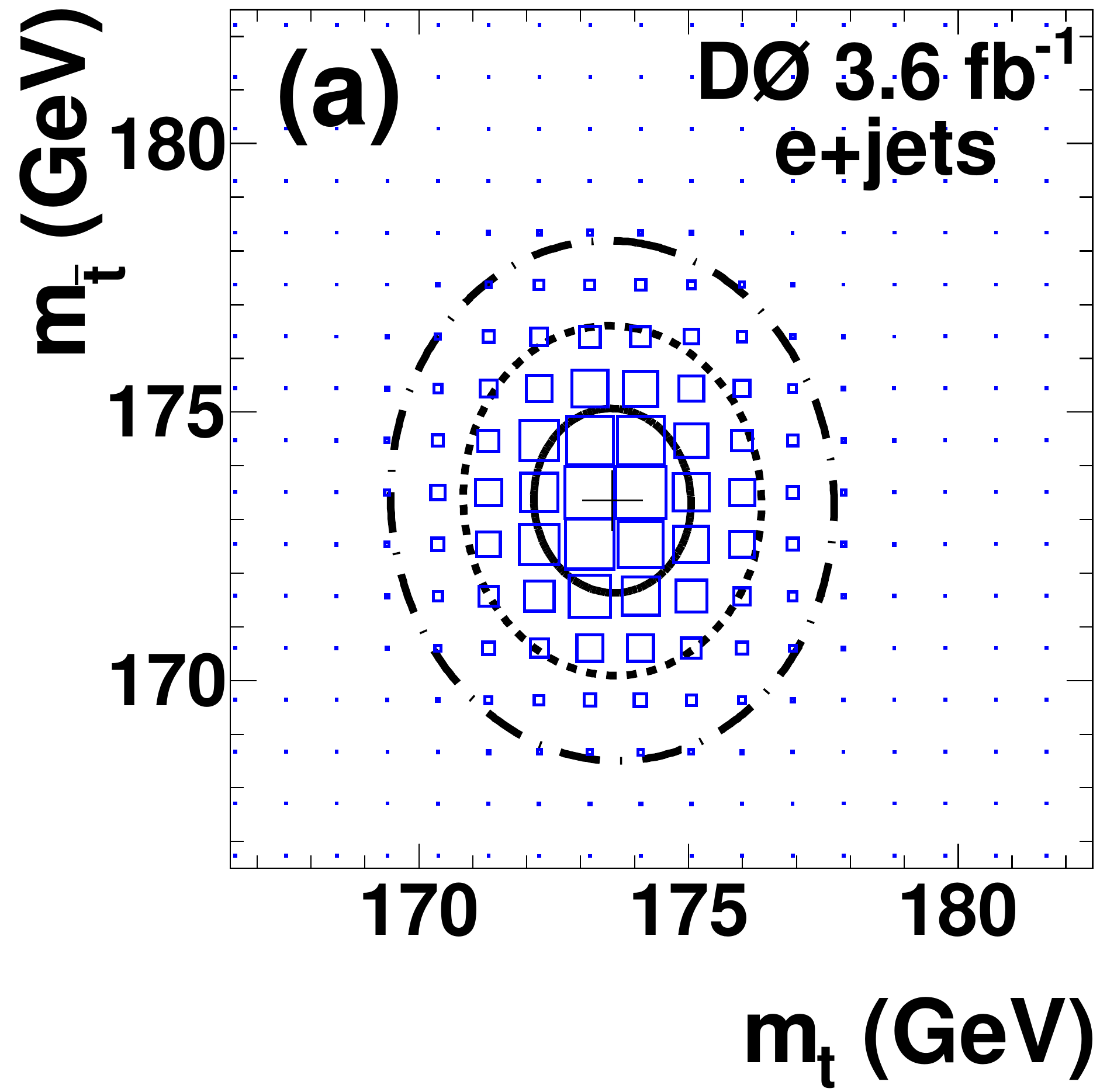}
\put(-1,2){(b)}
\end{overpic}
\qquad
\begin{overpic}[height=0.3\textwidth]{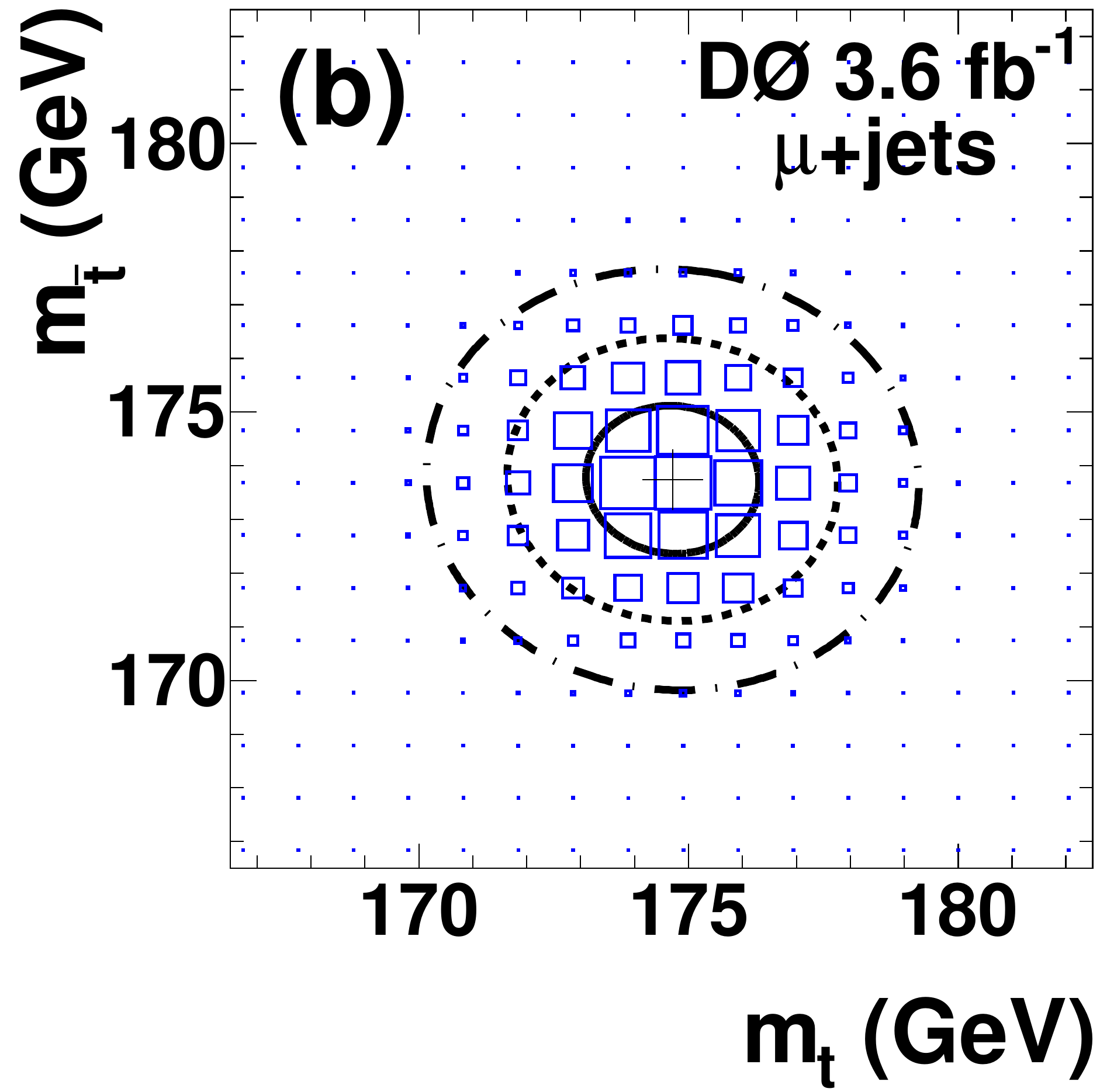}
\put(-1,2){(c)}
\end{overpic}
\caption{
\label{fig:mpole}
{\bf(a)} \sigtt\ measured by D0 using 5.3~\fb\ (black line) and theoretical NLO+NNLL~$^{16}$ (green solid line) and approximate NNLO~$^{17}$ (red solid line) predictions as a function of $\mpole$, assuming $\mmc=\mpole$. 
The gray band corresponds to the total uncertainty on measured $\sigtt$. The dashed lines indicate theoretical uncer\-tainties from the choice of scales and parton distribution functions. {\bf(b)} \mt\ and \mtb\ measured by D0 directly~and independently using 3.6~\fb\ in \ejets\ final states. The solid, dashed, and dash-dotted lines represent the 1, 2, and 3 SD contours. {\bf(c)} same as (b) but for \mujets.
}
\end{figure}

The invariance under $\mathcal{CPT}$ transformations is a fundamental property of the SM. $m_{\rm particle}\neq m_{\rm antiparticle}$ would constitute a violation of $\mathcal{CPT}$, and has been tested extensively in the charged lepton sector. Given its short decay time, the top quark offers a possiblity to test $\mt=\mtb$ at the \%-level, which is unique in the quark sector. D0 applies the ME technique to measure $\mt$ and $\mtb$ directly and independently using 3.6~\fb\ of data, and finds $\dm\equiv\mt-\mtb=0.8\pm1.8~\GeV$~\cite{bib:dm_d0}, in agreement with the SM prediction. The results are illustrated in Fig.~\ref{fig:mpole}. With 0.5~\GeV, the systematic uncertainty on \dm\ is much smaller than that on $\mtop$ due to cancellations in the difference, and is dominated by the uncertainty on the difference in calorimeter response to $b$ and $\bar b$ quark jets.

\vspace{-3mm}

\section{Tevatron combination and outlook}
Currently, the world's most precise measurements of \mtop\ are performed by CDF and D0 collaborations in \ljets\ final states. The preliminary Tevatron combination using up to 5.8 fb$^{-1}$ of data results in $m_{\rm top} = 173.2 \pm 0.9$~GeV~\cite{bib:combi}, corresponding to a relative uncertainty of 0.54\%. More details on the Tevatron combination can be found in Ref.~\cite{bib:combitalk}. 

With about 10.7~\fb\ recorded, the precision on \mtop\ is expected to further improve at D0, since only 3.6~\fb\ are used in the flagship measurement in \ljets\ final states. This applies not only to the statistical uncertainty, but also to several systematic uncertainties due to the limited size of calibration samples, like e.g.\ some components of the JES. Moreover, efforts are under way to better understand systematic uncertainties from the modelling of \ttbar\ signal, in particular the dominating uncertainty from different hadronisation and underlying event models. We look forward to exciting updates of \mtop\ measurements presented here.


\section*{Acknowledgments}
I would like to thank my collaborators from the D0 experiment for their help in preparing this article. I also thank the staffs at Fermilab and collaborating institutions, as well as the D0 funding agencies.



\begin{thebibliography}{99}

\bibitem{bib:topdiscovery}
F. Abe \etal\ (CDF Coll.), Phys. Rev. Lett. {\bf 74}, 2626 (1995), 
S.~Abachi~\etal\ (D0 Coll.), Phys. Rev. Lett. {\bf 74}, 2632 (1995).

\bibitem{bib:singletop}
T. Aaltonen \etal\ (CDF Coll.), Phys. Rev. Lett. {\bf 103}, 092001 (2009),
V.~M.~Abazov~\etal\ (D0 Coll.), Phys. Rev. Lett. {\bf 103}, 092002 (2009).

\bibitem{bib:wmasstalk}
V. M. Abazov \etal\ (D0 Coll.), Phys. Rev. Lett. {\bf 108}, 151804 (2012).

\bibitem{bib:proptalk}
C. Schwanenberger, these proceedings; Y. Peters, these proceedings.

\bibitem{bib:topresd0}
\verb|http://www-d0.fnal.gov/Run2Physics/WWW/results/top.htm|,
\verb|http://www-d0.fnal.gov/Run2Physics/WWW/documents/Run2Results.htm|.

\bibitem{bib:xsec}
C. Schwanenberger, these proceedings.

\bibitem{bib:singletoptalk}
Y. Peters, these proceedings.

\bibitem{bib:topmassd0nature}
V. M. Abazov \etal\ (D0 Collaboration),
Nature \textbf{429}, 638 (2004).

\bibitem{bib:mtoplj_d0}
V. M. Abazov \etal\ (D0 Coll.), Phys. Rev. D {\bf84}, 032004 (2011).

\bibitem{bib:mtopllme_d0}
V. M. Abazov \etal\ (D0 Coll.), Phys. Rev. Lett. 107, 082004 (2011).

\bibitem{bib:mtopll_d0}
V. M. Abazov \etal\ (D0 Coll.), Fermilab-Pub-12/020-E, submitted to Phys.~Rev.~Lett., arXiv:1201.5172 [hep-ex] (2012).

\bibitem{bib:mtoppole_d0}
V. M. Abazov \etal\ (D0 Coll.), Phys. Lett. B {\bf 703}, 422 (2011).

\bibitem{bib:signlo}
V.~Ahrens \etal, J. High Energy Phys. {\bf1009} (2010) 097, Nucl. Phys. B (Proc. Suppl.) 205-206 (2010) 48.

\bibitem{bib:signnlo}
S.~Moch, P.~Uwer, Phys. Rev. D {\bf78} (2008) 034003; U.~Langenfeld, S.~Moch, P.~Uwer, Phys. Rev. D {\bf80} (2009) 054009.

\bibitem{bib:dm_d0}
V. M. Abazov \etal\ (D0 Coll.), Phys. Rev. D {\bf 84}, 052005 (2011).

\bibitem{bib:combi}
The Tevatron Electroweak Working Group and CDF and D0 Collaborations, arXiv:1007.5255 [hep-ex] (2011).

\bibitem{bib:combitalk}
F. D\'eliot, these proceedings.

\end{thebibliography}
\end{document}